\documentclass[aps,prb,amsmath,twocolumn,superscriptaddress]{revtex4}

\usepackage{amsmath}
\usepackage{amssymb}

\usepackage{graphicx}
\usepackage{bm} 
\usepackage{url}

\usepackage{color}

\usepackage{amsmath}
\DeclareMathOperator{\Tr}{Tr}
\DeclareMathOperator{\Imag}{Im}
\DeclareMathOperator{\Real}{Re}

\begin{document}

\title{Interplay between Josephson and Aharonov-Bohm effects in Andreev interferometers}
\author{Pavel E. Dolgirev}
\affiliation{Skolkovo Institute of Science and Technology, Skolkovo Innovation Center, 3 Nobel St., 143026 Moscow, Russia}
\author{Mikhail S. Kalenkov}
\affiliation{I.E. Tamm Department of Theoretical Physics, P.N. Lebedev Physical Institute, 119991 Moscow, Russia}
\affiliation{Moscow Institute of Physics and Technology, Dolgoprudny, 141700 Moscow region, Russia}

\author{Andrei D. Zaikin}
\affiliation{Institut f{\"u}r Nanotechnologie, Karlsruher Institut f{\"u}r Technologie (KIT), 76021 Karlsruhe, Germany}
\affiliation{National Research University Higher School of Economics, 101000 Moscow, Russia}

\date{\today}

\begin{abstract}
Proximity induced quantum coherence of electrons in multi-terminal voltage-driven hybrid normal-superconducting nanostructures  may result in a non-trivial interplay between topology-dependent Josephson and Aharonov-Bohm effects. We elucidate a trade-off between stimulation of the voltage-dependent Josephson current due to non-equilibrium effects and quantum dephasing of quasiparticles causing reduction of both Josephson and Aharonov-Bohm currents. We also predict phase-shifted quantum coherent oscillations of the induced electrostatic potential as a function of the externally applied magnetic flux. Our results may be employed for engineering superconducting nanocircuits with controlled quantum properties.
\end{abstract}

\maketitle

\section{Introduction}

Long-range quantum coherence under non-equilibrium conditions in normal-superconducting (NS) heterostructures manifests itself in a large number of interesting and non-trivial phenomena \cite{belzig1999quasiclassical}. These phenomena become particularly pronounced in the low temperature limit since in this case proximity-induced quantum coherence of electrons in a normal metal may persist
even far away from a superconductor being limited only by dephasing due to electron-electron interactions \cite{deph1,deph2}.

In multi-terminal hybrid NS nanostructures (also called Andreev interferometers) one can easily drive electrons out of equilibrium by applying an external voltage bias to (some of) the normal terminals. In three-terminal NSN systems long-range quantum coherence of electrons results in conductance anomalies associated with non-local Andreev reflection \cite{Pino,BeckmannCAR,TeunCAR,VenkatCAR,KZCAR,GKZ}. Non-trivial phenomena also occur in  cross-like structures
 with two normal and two superconducting terminals interconnected by normal wires~\cite{V,WSZ,Yip,Teun}. Biasing the normal terminals by some voltage $V$, one can control both the magnitude and the phase dependence of the supercurrent between the two superconducting terminals  demonstrating switching between 0- and $\pi$-junction states \cite{V,WSZ,Yip,Teun}. In other words, in this case the dc Josephson current $I_J$ between the two S-terminals is determined not only by the superconducting phase difference $\phi$ but also by the bias voltage $V$, i.e. $I_J=I_J(V,\phi)$.

Likewise, dissipative currents in multi-terminal hybrid superconducting circuits can also be controlled both by external voltage and
the superconducting phase \cite{nakano1991quasiparticle} further emphasizing a non-trivial interplay between quantum coherence and non-equilibrium effects.
In ring-shaped geometries one can conveniently fix the phase difference $\phi$ by inserting an external magnetic flux $\Phi$ inside the
ring and investigate proximity-enhanced Aharonov-Bohm current oscillations \cite{Petr,Zaitsev,SN96,Grenoble,GWZ97} $I_{AB}(V,\phi)$, where $\phi=2\pi \Phi/\Phi_0$ and $\Phi_0$ is the superconducting flux quantum.

Thus, in NS hybrid nanostructures there exist two physically different contributions to the current -- $I_J(V,\phi)$ and $I_{AB}(V,\phi)$ --
sensitive to both proximity-induced quantum coherence and non-equilibrium conditions. Until recently these two currents had been investigated
separately from each other. For instance, no Josephson current can possibly occur in ring-shaped NS structures \cite{Grenoble,SN96,GWZ97}
where Aharonov-Bohm oscillations of the current $I_{AB}(V,\phi)$ have been demonstrated \cite{Grenoble}. And vice versa, no
Aharonov-Bohm effect can emerge in {\it symmetric} cross-like four-terminal setups \cite{WSZ,Yip,Teun} where the voltage-controlled dc Josephson current
has been observed \cite{Teun}.

Recently we argued \cite{PD18} that by slightly modifying the topology of a four-terminal Andreev interferometer -- e.g., just by making the cross-like
geometry \cite{WSZ,Yip,Teun} {\it asymmetric} -- one can induce non-vanishing Aharonov-Bohm currents, thus being able to directly observe a trade-off between Josephson and Aharonov-Bohm effects in the same setup. The competition between the two $2\pi$-periodic in $\phi$ terms $I_J(V,\phi)$ and $I_{AB}(V,\phi)$ -- respectively odd and even functions of $\phi$ -- yields novel features such as, e.g., the $(I_0,\phi_0)$-junction state \cite{PD18} for which the current-phase relation turns out to be
phase-shifted by the value $\phi_0$ controlled by an external voltage bias $V$. Interestingly enough, applying a temperature gradient to the system one can induce the thermoelectric voltage signal which
also demonstrates coherent phase-shifted oscillations as a function of $\Phi$. Such oscillations turn out to be quite similar \cite{PD18,PD18Th} (although not exactly identical) to those of the electric current.

In this work we will further investigate a non-trivial interplay between dissipative (Aharonov-Bohm) and
non-dissipative (Josephson) contributions to the current in multiterminal Andreev interferometers at low temperatures and under non-equilibrium conditions. In particular, we will demonstrate that providing extra low energy quasiparticles in a voltage biased setup (e.g., by attaching an extra normal terminal to the system) yields a trade-off between effective dephasing of quasiparticles (causing reduction of both Josephson and Aharonov-Bohm currents) and stimulation of the (voltage-dependent) Josephson current. The combination of these two effects may result in a substantial modification of the current-phase relation in Andreev interferometers and to further interesting topology-dependent phenomena like, e.g., coherent oscillations of the voltage induced at the normal terminal isolated from external leads.

The structure of our paper is as follows. In Sec. II we define our model and briefly
describe the quasiclassical formalism employed in our calculations. In Sec. III we discuss phase-shifted coherent oscillations of the current in four-terminal Andreev interferometers. Sec. IV is devoted to the analysis of novel non-equilibrium effects which occur in the five-terminal configurations. In Sec.~\ref{sec: Conclusions}, we briefly summarize our main findings and provide further discussion. Some details of our calculation are relegated to Appendix.

\section{The model and basic formalism}
\label{sec: Formalism}

\begin{figure}
\centering
\includegraphics[width=0.8\linewidth]{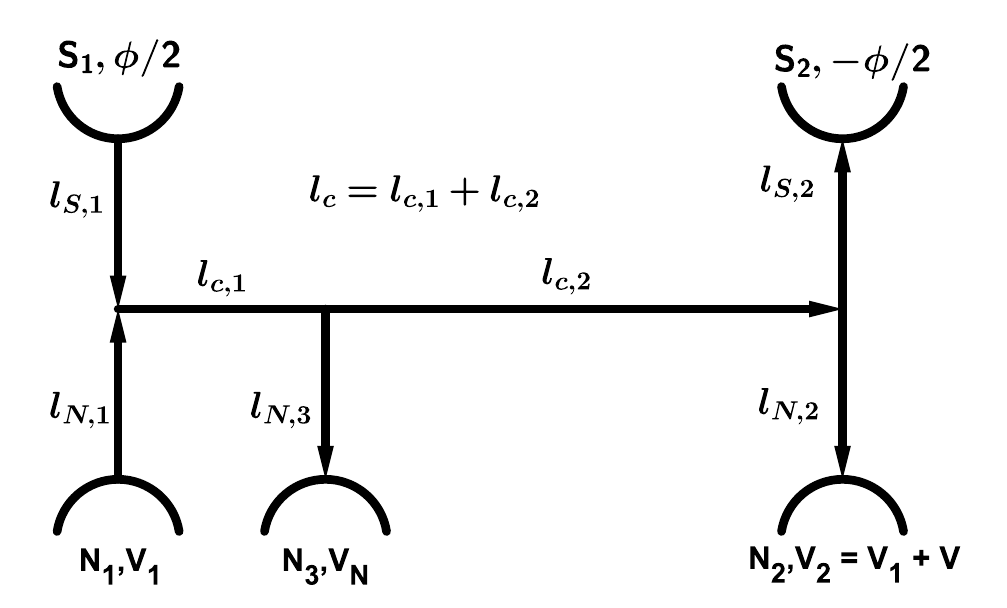}
\caption{Schematics of a five-terminal Andreev interferometer under consideration. It consists of two superconducting (S$_{1,2}$) and three normal (N$_{1,2,3}$) terminals interconnected by normal metallic wires of lengths $l_{c,1,2}$, $l_{N,1,2}$ and $l_{S,1,2}$. The two normal terminals N$_1$ and N$_2$ are biased by a constant voltage $V=V_2 - V_1$, while the third normal terminal N$_3$ remains isolated from any external circuit.
The phase difference $\phi$ between the two superconducting terminals can be controlled by an external magnetic
flux inside the loop formed by these terminals. Arrows indicate the (chosen as a convention) directions of the current flowing inside the corresponding wires.}
\label{fig: geom}
\end{figure}

Below in this work we will mainly focus our attention on a five-terminal hybrid NS structure consisting of two superconducting (S$_1$ and S$_2$) and three normal (N$_{1,2,3}$) terminals interconnected by normal diffusive wires of equal cross section ${\cal A}$ and different lengths as it is illustrated in Fig.~\ref{fig: geom}.
The superconducting order parameter in the two S-terminals has the form $\Delta e^{\pm i\phi/2}$, implying that the phase difference between these terminals equals to $\phi$. The two normal terminals N$_1$ and N$_2$ are attached to an external voltage source thus fixing the voltage difference between these terminals $V_2-V_1=V$. The third terminal N$_3$ is kept isolated from any external circuit. Nevertheless, depending on the system topology a non-zero electric potential $V_N=V_N(V,\phi)$ may be generated at this terminal.

In what follows we will assume that the effective distance between the two superconducting terminals $L = l_{S,1} + l_c + l_{S,2}$ strongly exceeds the superconducting coherence length $\xi$. Then the corresponding Thouless energy of our device ${\mathcal E}_{\rm Th} = D/L^2$ (with $D$ being the wire diffusion constant) remains  well below the superconducting gap, i.e. we have ${\mathcal E}_{\rm Th} \ll \Delta$.

Our further analysis will be based on the well established quasiclassical formalism of the superconductivity theory~\cite{belzig1999quasiclassical}. Employing the so-called $\theta$-parameterisation \cite{ZZh,belzig1999quasiclassical} we express the $2\times 2$ matrix in the Nambu space representing the retarded quasiclassical Green function $\hat{G}^R$ in the form
\begin{equation}
\hat{G}^R = \begin{pmatrix}
{\cal G}_{11} & {\cal F}_{12}\\
{\cal F}_{21} & {\cal G}_{22}
\end{pmatrix} = \begin{pmatrix}
\cosh \theta & e^{i\chi} \sinh \theta \\
-e^{-i\chi} \sinh\theta  & -\cosh\theta
\end{pmatrix},
\end{equation}
where $\theta$ and $\chi$ are two complex functions obeying the spectral Usadel equations
\begin{eqnarray}
D \Delta \theta & = & -2 i \epsilon \sinh \theta + \frac{1}{2} D (\nabla \chi)^2 \sinh 2\theta, \label{eq: Theta}\\
\nabla j_E & = & 0,\ j_E = \sinh^2 \theta \cdot \nabla \chi.
\label{eq: chi}
\end{eqnarray}

The quantum kinetic equations read \cite{belzig1999quasiclassical}
\begin{align}
& \nabla j_L = 0 ,\ j_L =  D_L \nabla f_L - \mathcal{Y} \nabla f_T + j_s f_T, \label{eq: j_L}\\
& \nabla j_T = 0 ,\ j_T =  D_T \nabla f_T + \mathcal{Y}\nabla f_L + j_s f_L,\label{eq: j_T}
\end{align}
where $f_{L(T)}(\epsilon)$ is symmetric (antisymmetric) in energy part of the electron distribution function.
In Eqs.~(\ref{eq: j_L})--(\ref{eq: j_T}) we also introduced the kinetic coefficients
\begin{align}
&D_{L/T} = \frac{1}{2} (1 +  | \cosh \theta |^2 \mp |\sinh \theta|^2 \cosh ( 2\Imag{\chi} )),\\
&\ \mathcal{Y} = \frac{1}{2}|\sinh \theta|^2\sinh ( 2\Imag{\chi} ),\  j_s = \Imag{j_E}.
\end{align}
It is worth pointing out that the function ${\cal Y}$ accounts for electron-hole asymmetry in our structure.

The electric current density $j$ in our system is expressed in terms of the energy-integrated $j_T$-component of the spectral current as
\begin{equation}
j = -\frac{\sigma_N}{2e} \int j_T (\epsilon) d \epsilon ,\label{current}
\end{equation}
where $\sigma_N$ is the Drude conductivity of a normal metal. By solving the above Usadel equations one can also determine the distribution of the electrostatic potential in our structure by means of the formula
\begin{equation}
e V(x)=  \int_0^\infty d \epsilon f_T (x,\epsilon) \nu_\epsilon (x) ,
\label{potential}
\end{equation}
where $\nu_\epsilon (x) = \Real \{\cosh \theta (x,\epsilon)\}$ is the coordinate-dependent electron density of states.

Eqs.~(\ref{eq: Theta})--(\ref{eq: j_T}) should be solved separately in each of the metals and the corresponding solutions should be matched with the aid of proper boundary conditions at all interfaces of our structure. Here we will assume that all inter-metallic interfaces are fully transparent implying that at all wire nodes (a) the functions $\theta,\ \chi,\ f_L$ and $f_T$ are continuous and (b) the spectral currents ${\cal A}\nabla \theta  ,\ {\cal A}  \nabla \chi, \ {\cal A} j_L$ and ${\cal A} j_T$ remain conserved. In addition, at the boundaries between the wires and the N-terminals the functions $\theta$, $\chi$, $f_L$ and $f_T$ are continuously matched with their bulk values deep inside these terminals. At the NS interfaces we have $j_L = 0$ and $f_T = 0$ at energies $| \epsilon | < \Delta$. The latter condition just means that charge imbalance (possibly existing inside normal wires) disappears at the NS interfaces.

\section{Four-terminal interferometer}

To begin with, let us somewhat simplify our system and disconnect the N$_3$ terminal from the rest of the structure, thus reducing our five-terminal system depicted in Fig.~\ref{fig: geom} to a four-terminal one \cite{PD18}. It is instructive to first discuss the behavior of this reduced structure
since it will help us to elucidate all essential physics and to make our subsequent analysis of the five-terminal setup of Fig.~\ref{fig: geom} a lot easier. In addition, for simplicity we set $l_{S(N),1} = l_{S(N), 2} = l_{S(N)}$, in which case one has $V_{1/2} = \mp V/2$.

As usually, in order to proceed, one first solves the spectral part of the problem and finds the retarded and advanced Green functions for the structure under consideration. This task can easily be accomplished: At energies $|\epsilon| \ll \mathcal{E}_{\rm Th}$ and $|\epsilon| \gg \mathcal{E}_{\rm Th}$ an analytic solution can be obtained~\cite{ZZh} (see also Appendix), while for $|\epsilon| \sim \mathcal{E}_{\rm Th}$ it is in general necessary to resort to numerics~\cite{GreKa,VH}.

The next step is to resolve the kinetic equations. The corresponding solution can also be obtained analytically
provided we \cite{PD18} (i) disregard terms containing $\mathcal{Y}$ and (ii) resolve the kinetic equations in the first order in $j_s$. Strictly speaking, the approximations (i) and (ii) are fully justified only in the vicinity of the phase values $\phi \approx \pi n$. Fortunately, the exact numerical analysis of the problem~\cite{PD18} verifies that the above approximations work sufficiently well allowing to capture all essential physics even far away from $\phi \approx \pi n$.

Assuming that the superconducting order parameter $\Delta$ strongly exceeds any other energy scale in our problem, in the leading order in $j_s$ we find
\begin{eqnarray}
&f_L(x) \equiv f_L^N (V/2),\\
&f_{L/T}^N (V) = \frac{1}{2} \Big[ \tanh \frac{\epsilon + eV}{2T} \pm  \tanh \frac{\epsilon - eV }{2T}\Big],
\end{eqnarray}
and $j_L = 0$ in every wire of the structure.

Turning now to electric currents flowing in our system, for the wire $l_{N,1}$ we may write
\begin{equation}
j^N_T = D_T f_T' \Rightarrow j_T^N = (f_T^{c,1} - f_T^N(V_1)) \left[\int_{l_N} dx/D_T\right]^{-1},
\end{equation}
where $f_T^{c,1}$ is evaluated at the crossing point $c_1$ of the wires $l_{S,1}, l_{N,1}$ and $l_c$.
Similarly, for the wires $l_{S,1}$ and $l_{c}$ we get:
\begin{eqnarray}
&j_T^S = j_s^S f_L^N (V/2) + f_T^{c,1} \displaystyle \left[\int_{l_S}dx/D_T\right]^{-1},\\
&j_T^c =  j_s^c f_L^N (V/2) - 2f_T^{c,1} \displaystyle \left[\int_{l_c} dx/D_T\right]^{-1}.
\end{eqnarray}
Making use of the conservation of the spectral charge current, $j_T^c  = j_T^S  + j_T^N $, we eventually recover the expression for the spectral current $I_S(\epsilon) = \sigma_N j_T^S {\cal A}/(2e)$, which, after energy integration, determines the current $I_S$ flowing between the two superconducting terminals S$_1$ and S$_2$:
\begin{equation}
I_S = \int d\epsilon \left[\sigma_N f_L^N (V/2)j_s {\cal A}/(2e) - f_T^N(V/2) {\cal R}_c^T/{\cal N}\right], \label{eqn: IS}
\end{equation}
where we denoted ${\cal N} = {\cal R}_c^T ({\cal R}_S^T + {\cal R}_N^T) + 2 {\cal R}_S^T {\cal R}_N^T$ and
${\cal R}^T_i = \displaystyle({\cal A}\sigma_N)^{-1} \int_{l_i} dx/D_{T,i}$.

Making use of Eq. (\ref{eqn: IS}) we obtain~\cite{PD18}
\begin{equation}
I_S = I_0(V) + I_J(V,\phi) + I_{AB}(V,\phi).
\end{equation}
The first term in the right-hand side of this formula represents the averaged over $\phi$ current value $I_0(V)=\langle I_S\rangle_\phi$, while two other -- sensitive to the phase -- terms are respectively the Josephson (odd in $\phi$) and the Aharonov-Bohm (even in $\phi$) contributions to the current. In the interesting for us limit of sufficiently large bias voltages $eV \gg {\mathcal E}_{\rm Th}$ we find~\cite{PD18}
\begin{equation}
I_J(V,\phi) \simeq I_C^{(4)}(V) \sin \phi, \quad  I_{AB}(V,\phi)\simeq I_{\rm m}^{(4)}(V) \cos \phi \label{IJIAB}
\end{equation}
with (see also Appendix)
\begin{eqnarray}
\label{eq: I_C}
&I_C^{(4)} (V) \simeq \frac{128 (1+v^{-1})}{9(3+2\sqrt{2})} \frac{V}{R_L} e^{-v} \sin(v + v^{-1}),\\
&I_{\rm m}^{(4)}\simeq \frac{0.18 \mathcal{E}_{\rm Th}}{e R_L}.
\label{eq: I_m}
\end{eqnarray}
In Eq. (\ref{eq: I_C})  we introduced the dimensionless parameter $v = \sqrt{\frac{eV}{2\mathcal{E}_{\rm Th}}} \gg 1$.

\begin{figure}
\begin{minipage}[h]{0.49\linewidth}
\center{\includegraphics[width=1\linewidth]{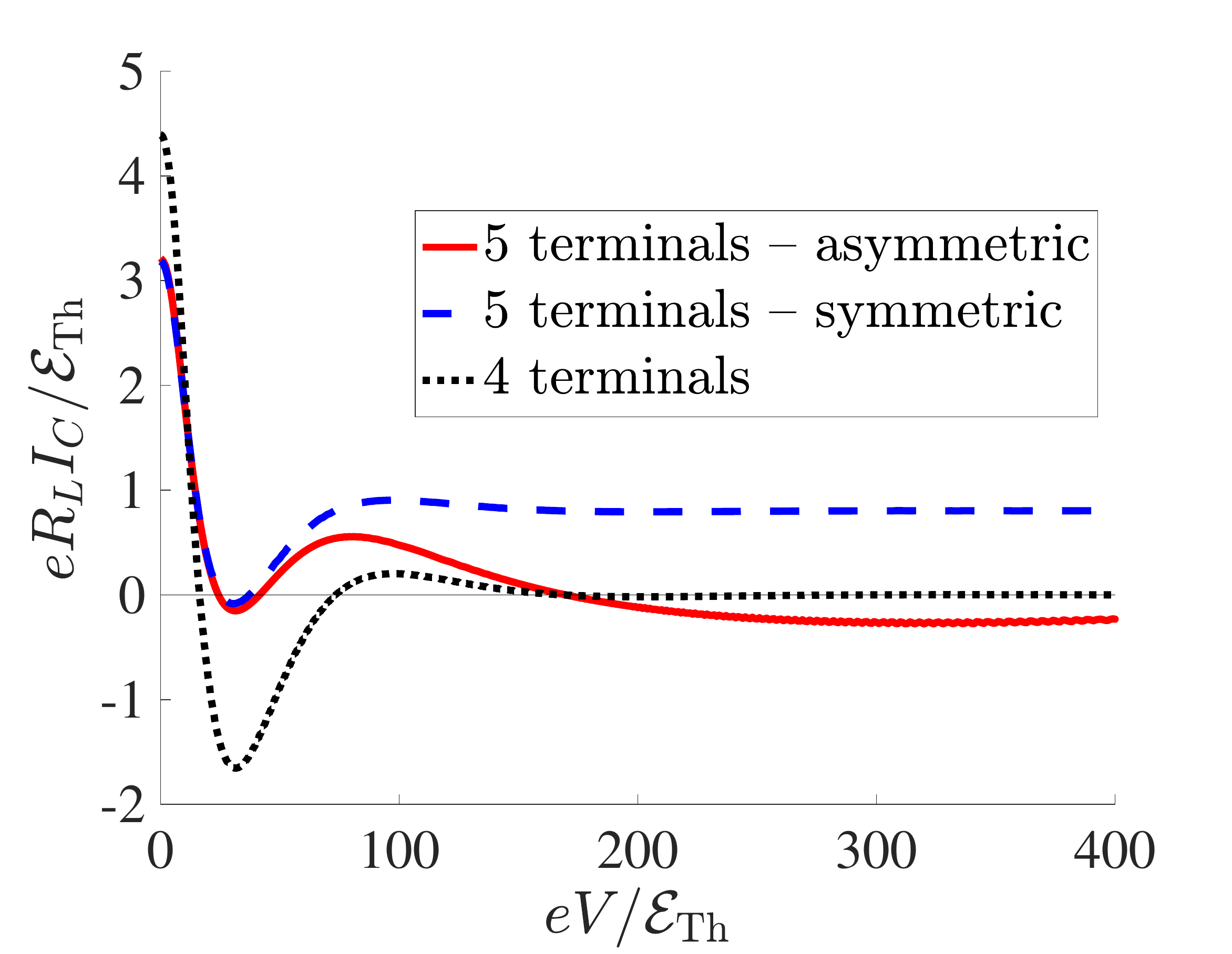}} \\a) \\
\end{minipage}
\hfill
\begin{minipage}[h]{0.49\linewidth}
\center{\includegraphics[width=1\linewidth]{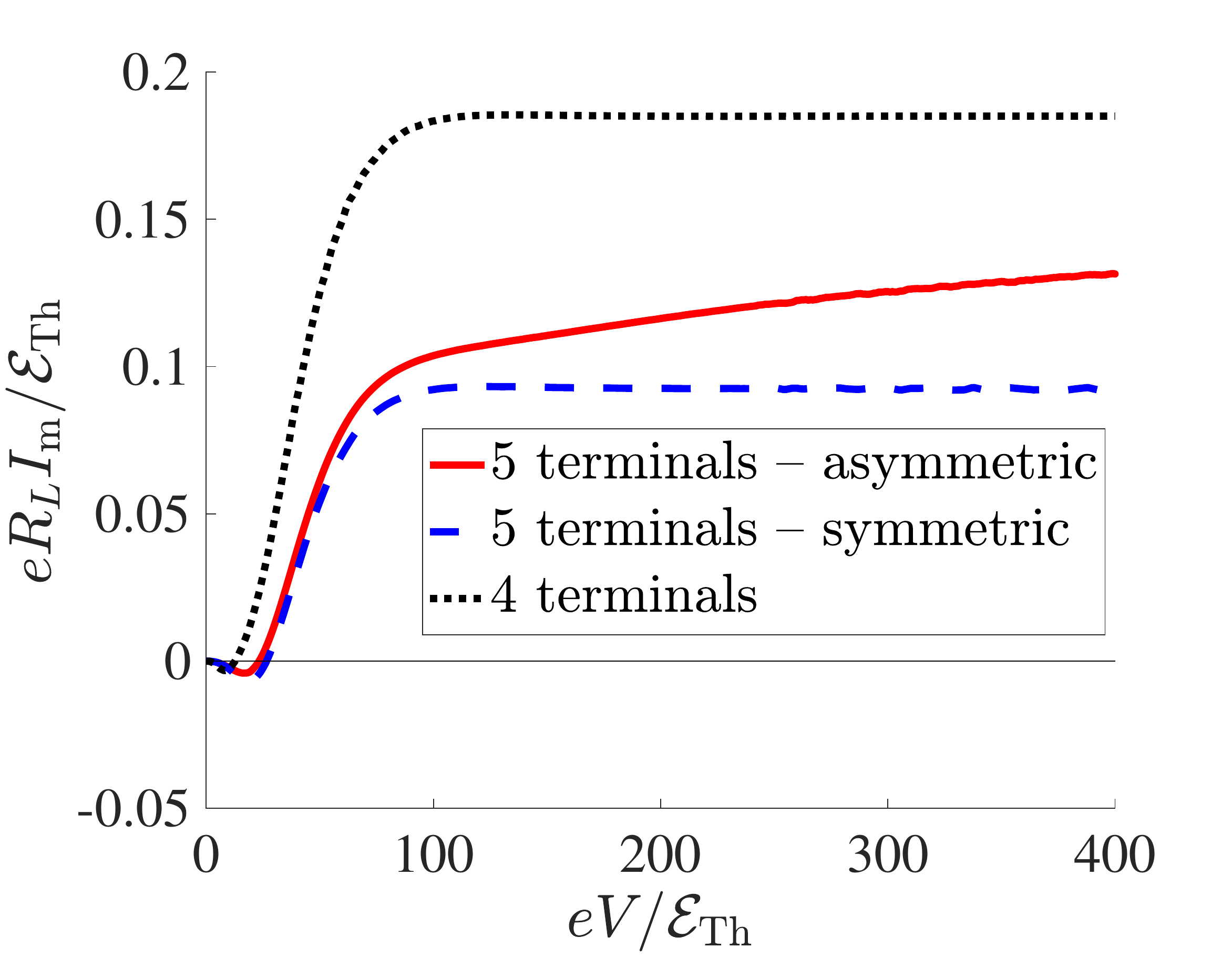}} \\b)
\end{minipage}
\caption{Left panel: Zero temperature Josephson critical current values $I_C^{(4)}(V)$ and $I_C^{(5)}(V)$ respectively for four- and five-terminal setups. Right panel: The same for the Aharonov-Bohm amplitudes $I_{\rm m}^{(4)}(V)$ and $I_{\rm m}^{(5)}(V)$. Here we choose ${\mathcal E}_{\rm Th} = 10^{-3}\Delta$, $l_{S,1} = l_{S,2} = l_{N,1} = l_{N,2} = l_c = 1/3L$, and $l_{N,3} = 1/2L$. In the case of an asymmetric five-terminal geometry $l_{c,1} = 0.1L$ and $l_{c,2} = (1/3-0.1)L$, cf. Fig.~\ref{fig: geom}.}
\label{fig: Currents}
\end{figure}

Note that both results (\ref{eq: I_C}) and (\ref{eq: I_m}) hold only in the low temperature limit. In particular, Eq. \eqref{eq: I_C} is valid for $T \ll \sqrt{e|V|\mathcal{E}_{\rm Th}}$, while low temperature asymptotics~\eqref{eq: I_m} is correct even for a wider temperature range.  The full voltage dependence for both $I_C^{(4)}$ and $I_{\rm m}^{(4)}$ is illustrated in Fig.~\ref{fig: Currents}. We observe that at low voltages the Josephson critical current $I_C^{(4)}$ shows the $\pi$-junction feature~\cite{WSZ,Yip,Teun} and  dominates over the Aharonov-Bohm contribution $I_{\rm m}^{(4)}$, whereas at high voltages $|I_C^{(4)}|$ decays exponentially with increasing $V$ in accordance with Eq. (\ref{eq: I_C}). The Aharonov-Bohm  current shows just the opposite trend: $I_m^{(4)}$ increases with $V$ and saturates to the value in Eq.~(\ref{eq: I_m}) at $eV \gtrsim 100 \mathcal{E}_{\rm Th}$.

For completeness, let us also point out that the currents $I_C$ and $I_{\rm m}$ are described by very different temperature dependencies: The Josephson term decays exponentially with increasing temperature being completely suppressed already at $T\simeq 20 \mathcal{E}_{\rm Th}$, while the Aharonov-Bohm current decays much slower, typically as a power-law \cite{GWZ97,Grenoble}.

\section{Five-terminal interferometer}

Let us now go back to our initial five-terminal configuration schematically depicted in Fig.~\ref{fig: geom}.
In other words, as compared to the situation considered in the previous section we now attach an extra reservoir of normal electrons N$_3$ to the central wire $l_c$. At the first glance, an immediate and
obvious consequence of this modification could only be a reduction of superconducting correlations in our system and, hence, partial suppression of both Josephson and Aharonov-Bohm contributions to the current $I_S(V)$. This is because a certain fraction of ``phase-coherent electrons'' propagating in the normal wires connecting the two S-terminals can now make a ``detour'' into N$_3$ being replaced by electrons from the latter terminal which carry no information about the phase $\phi$. As it is demonstrated in Appendix, at $T \to 0$ and $eV \gg D/l_c^2$ this decoherence mechanism yields a reduction of the Josephson critical current for the five-terminal setup as
\begin{equation}
I_{C,0}^{(5)} (V)=\frac{2}{3} I_C^{(4)} (V),
\label{IC05}
\end{equation}
where $I_C^{(4)} (V)$ is defined in Eq. (\ref{eq: I_C}).

Likewise, the Aharonov-Bohm current component of $I_S(V)$ in the five-terminal setup gets reduced as compared to
that in the four-terminal one. As it is illustrated in Fig.~\ref{fig: Currents}b, for the symmetric case (see below) we have:
\begin{equation}
I_{\rm m,0}^{(5)} (V)\approx \frac{1}{2} I_{\rm m}^{(4)}(V),
\label{Im5V}
\end{equation}
where $I_{\rm m}^{(4)}(V)$ is specified in Eq. (\ref{eq: I_m}).

In what follows, we will demonstrate that along with the above decoherence scenario, there is yet another effect which, on the contrary, may yield a significant {\it enhancement} of the Josephson current. On top of that, by applying an external voltage bias $V$,  we, in general, induce a non-zero electric potential $V_N$ at the terminal N$_3$. Below we will observe that the voltage $V_N(V, \phi)$ is also sensitive to proximity-induced quantum coherence effects and, hence, $V_N$ exhibits the (phase shifted) coherent oscillations as a function of the superconducting phase $\phi$.

\subsection{Symmetric setup}

We start by considering a fully symmetric configuration, in which case the terminal N$_3$ is connected by the wire $l_{N,3}$ to the central point of the wire $l_c$. As before, we also set $l_{S(N),1} = l_{S(N), 2} = l_{S(N)}$. Then by symmetry we have $V_N \equiv 0$ and $V_{1/2} \equiv \mp V/2$, i.e. no further evaluation of $V_N(V, \phi)$ would be necessary in this case. Adopting the same set of approximations and employing the same analysis as in the previous section, we evaluate the spectral Josephson current between the two superconducting terminals with the result
\begin{eqnarray}
I_J(\epsilon) &\simeq & [ (1-\kappa) f_L^N(V/2) + \kappa f_L^N(0) ] \frac{\sigma_N j_s {\cal A}}{2e},\label{eqn: I_J_5term} \\
\kappa &=& \frac{2 R_{N,1}}{R_c + 2 R_{N,1} + 4 R_{N,3}}.
\end{eqnarray}
The first term in the right-hand side of Eq.~(\ref{eqn: I_J_5term}) has exactly the same origin as the corresponding contribution in Eq.~(\ref{eqn: IS}) controlled by the voltage $V$ between the normal terminals N$_1$ and N$_2$. In contrast, the last term is new. It emerges here only due to the presence of the terminal N$_3$ not considered in the previous section. Since the voltage $V_N = 0$, the latter term turns out to be {\it independent} of the bias voltage $V$. Then, in the interesting limit $T \ll \sqrt{e |V| {\mathcal E}_{\rm Th}} \ll e|V|$ we obtain:
\begin{eqnarray}
I_J \simeq (1-\kappa) I_{C,0}^{(5)}(V) \sin \phi + \kappa I_{\rm eq}^{(5)}(\phi ),
\label{I5phi}
\end{eqnarray}
where $I_{\rm eq}^{(5)}(\phi )$ is the {\it equilibrium} Josephson current for the five-terminal setup of Fig.~\ref{fig: geom} at $T \to 0$. This current differs from that for an SNS junction~\cite{ZZh,GreKa} only by a geometry-dependent numerical prefactor smaller than unity.

Equations (\ref{eqn: I_J_5term})--(\ref{I5phi}) represent an important result: We observe that, while the first -- voltage controlled -- term in the right-hand side of Eq.~(\ref{I5phi}) decays exponentially with increasing $eV \gg {\mathcal E}_{\rm Th}$, the second term remains nonzero being equal to a voltage-independent constant, cf. also Fig.~\ref{fig: Currents}a. In other words, under these non-equilibrium conditions the maximum value of the Josephson current
\begin{equation}
I_{C}^{(5)}(V) \simeq \kappa \  {\rm max}_\phi [I_{\rm eq}^{(5)}(\phi )] \simeq 3.2 \kappa {\mathcal E}_{\rm Th}/(e R_L)
\end{equation}
may strongly exceed $I_{C,0}^{(5)}(V)$ in Eq.~(\ref{IC05}). The physical reason for this enhancement effect is transparent: The terminal N$_3$ supplies extra quasiparticles with energies $|\epsilon |\sim T$ --  well below both $eV$ and
${\mathcal E}_{\rm Th}$ -- to the wires connecting the two superconducting terminals. Accordingly, the Josephson
current acquires an extra contribution, which is not exponentially suppressed at low enough temperatures no matter how large the external bias $V$ is. Nevertheless, this non-equilibrium effect may be considered curious because the supercurrent enhancement is provided by the normal terminal N$_3$, which ``knows nothing'' about superconductivity at all.

\begin{figure}
\begin{minipage}[h]{0.49\linewidth}
\center{\includegraphics[width=1\linewidth]{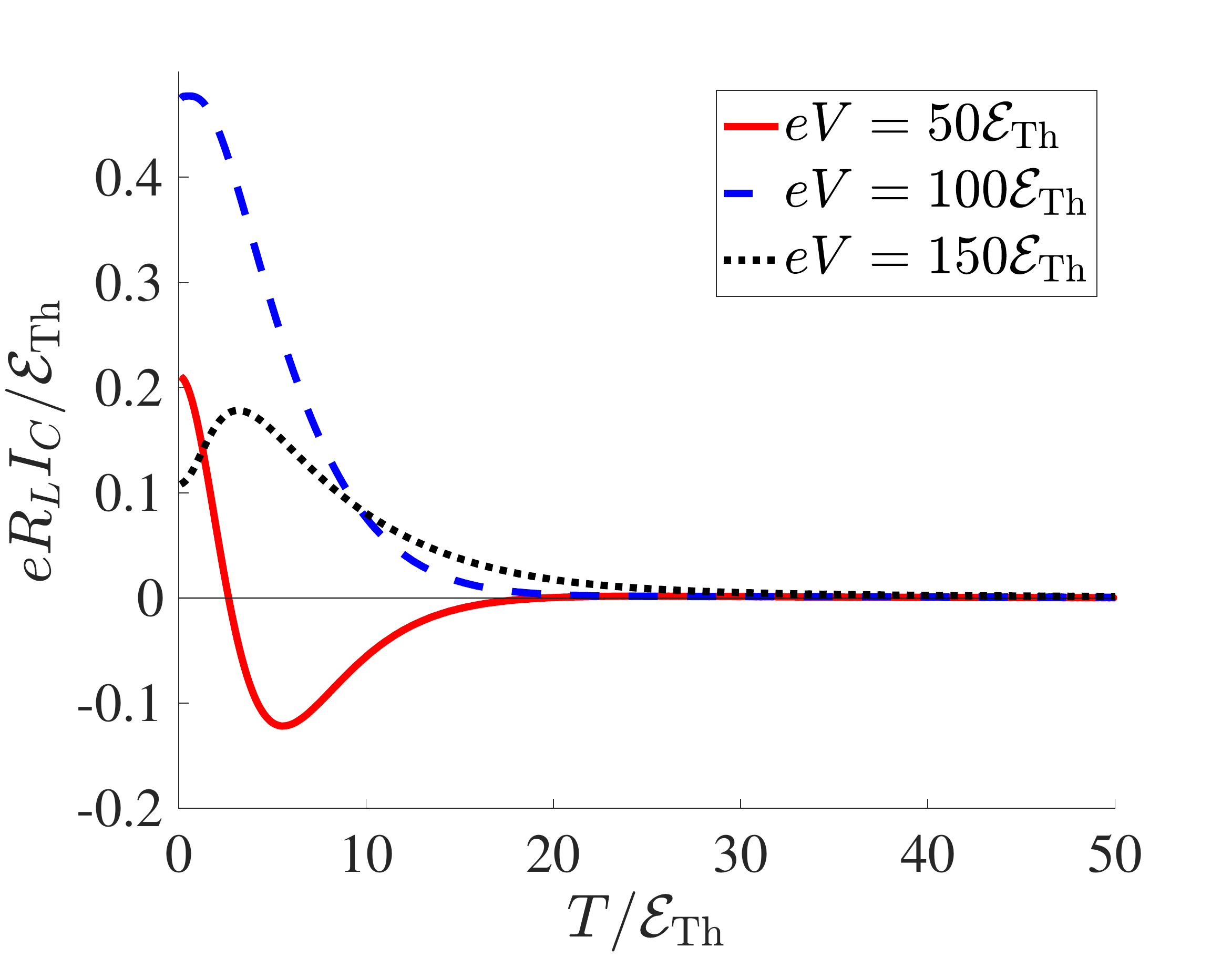}} \\a) \\
\end{minipage}
\hfill
\begin{minipage}[h]{0.49\linewidth}
\center{\includegraphics[width=1\linewidth]{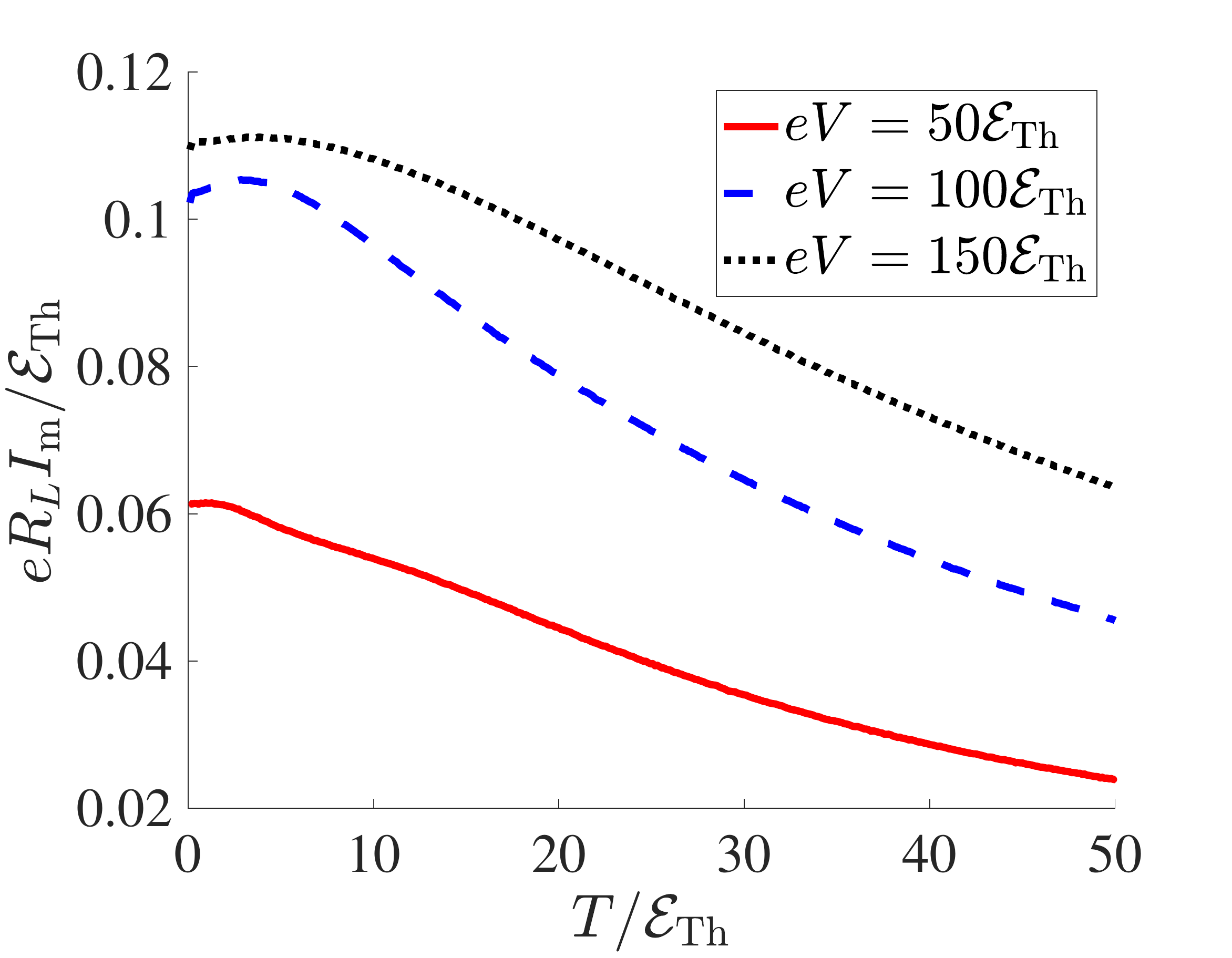}} \\b)
\end{minipage}
\caption{(a) The Josephson current amplitude $I_C^{(5)}$ for an asymmetric five-terminal setup of Fig.~\ref{fig: geom} as a function of temperature at different bias voltages $V$. (b) The same for the maximum Aharonov-Bohm current $I_{\rm m}^{(5)}$. The system parameters are the same as in Fig.~\ref{fig: Currents}.}
\label{fig: CMP_T}
\end{figure}

It is also interesting that, unlike for $I_J$, no such enhancement effect is observed for the Aharonov-Bohm contribution $I_{AB}$, here the only effect of the terminal N$_3$ is the current suppression (\ref{Im5V}), see also Fig.~\ref{fig: Currents}b. This tendency is also understandable since, unlike in the case of the supercurrent, low energy quasiparticles mainly contribute to the Aharonov-Bohm current even at high voltages \cite{GWZ97, Grenoble}. Accordingly, no significant impact of the terminal N$_3$ on $I_{AB}$ (apart from that accounted for by Eq. (\ref{Im5V})) could be expected. We can also add that with increasing temperature above
${\mathcal E}_{\rm Th}$ both current components $I_{C}^{(5)}$ and $I_{\rm m}^{(5)}$ decay (respectively exponentially and as a power-law) similarly to the case of a four-terminal setup. This behavior is illustrated in Fig.~\ref{fig: CMP_T} for the case of an asymmetric setup to be addressed below.

\subsection{Asymmetric setup}

\begin{figure}
\centering
\includegraphics[width=1\linewidth]{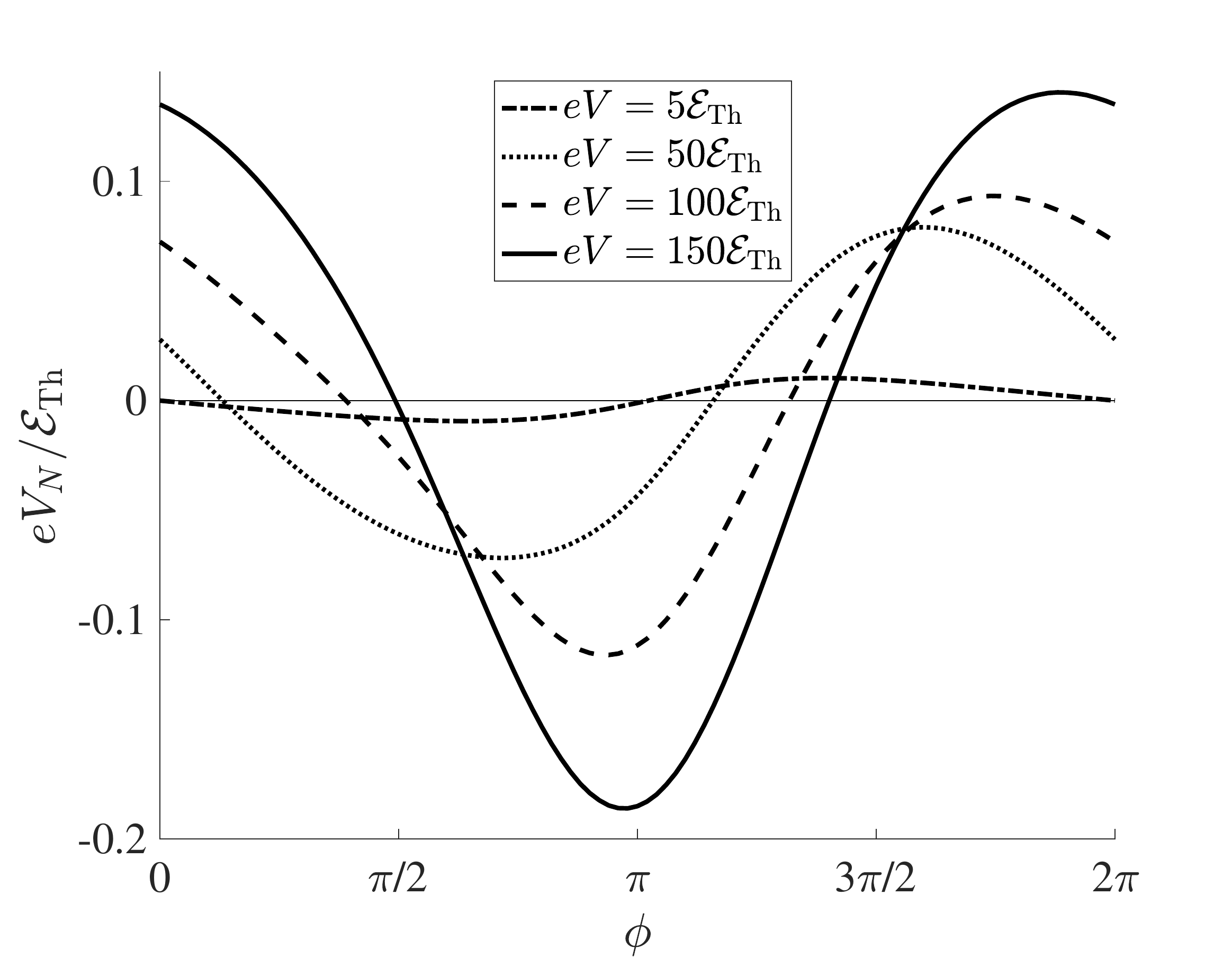}
\caption{An oscillating part of the induced voltage $V_N$ as a function of $\phi$ at $T \to 0$ and different bias voltages $V$. In the low-voltage limit the periodic function $V_N(\phi)$ is odd being converted into an even one at higher voltages.}
\label{fig: V_N}
\end{figure}

Let us now consider an asymmetic setup, in which case the terminal $N_3$ is attached to the wire $l_c$ in a non-symmetric fashion, just as it is shown in Fig.~\ref{fig: geom}. Then the problem gets somewhat more involved since the conditions $V_{1/2} = \mp V/2$ and $V_N = 0$ no longer apply. In other words, the voltages $V_{1/2}$ and $V_N$ should now be evaluated self-consistently by solving the Usadel equations combined with Eq.~(\ref{potential}). Treating this problem numerically, bearing in mind that ($i$) no current can flow into the normal terminal N$_3$, ($ii$) $I_{N,1} = I_{N,2}$ and ($iii$) $V= V_2 - V_1$, we arrive at the results for $V_N$ which contains an oscillating in $\phi$ part displayed in Fig.~\ref{fig: V_N}.

We observe that this oscillating part of the voltage $V_N$ depends on both $\phi$ and $V$: it is an odd-like $2\pi$-periodic function of the phase at smaller voltages $eV \lesssim 20{\cal E}_{\rm Th}$ and shows an even-like behavior at higher voltage values $eV \gtrsim 80{\cal E}_{\rm Th}$. Thus, the value $V_N(\phi, V)$ turns out to be sensitive to the proximity-induced long-range quantum coherence of the electrons in normal wires, and it essentially originates from an interplay between the Aharonov-Bohm and Josephson effects.

Without loss of generality the function $V_N(\phi, V)$ can be decomposed into even and odd terms as
\begin{equation}
V_N(\phi, V) = V_{\rm even}(\phi, V) + V_{\rm odd}(\phi, V).
\end{equation}
In order to estimate the even part, one can solve the kinetic equations analytically by setting $D_{L/T} \approx 1$ and neglecting both $j_s$ and ${\cal Y}$. Then one finds:
\begin{equation}
V_{\rm even} \approx \dfrac{R_{c_1} - R_{c_2}}{8(R_{c_1} + R_{c_2})} V + V_{\rm AB},
\end{equation}
where $R_{c_i}= l_i/(\sigma_N\mathcal{A})$ is the normal state resistance of the wire segment of length $l_i$.  Hence, the even in $\phi$ part of the voltage $V_N$ equals to the sum of Ohmic and Aharonov-Bohm terms, where at large enough $V$ the latter saturates to the value
\begin{equation}
V_{\rm AB} \approx 0.29 ({\cal E}_{\rm Th}/e)\cos\phi.
\end{equation}

As far as the odd in $\phi$ term $V_{\rm odd}$ is concerned, our numerical analysis demonstrates that, being important at smaller voltage values $V$, this term becomes strongly suppressed in the large voltage limit. This behavior is reminiscent of that for the currents $I_C^{(4)}(V)$ and $I_{C,0}^{(5)}(V)$, thereby indicating that the presence of the odd in $\phi$ contribution $V_{\rm odd}$ may be associated with the Josephson-like effect. At the same time, one should keep in mind that in the asymmetric setup one has ${\cal Y} \neq 0$ in the wire $l_{N,3}$. Hence, electron-hole asymmetry~\cite{KZ17} induced in the kinetic equations by the ${\cal Y}$-term  should also be taken into account while evaluating the contribution $V_{\rm odd}$. More detailed description of the electron-hole asymmetry effects is beyond the scope of the present paper and will be presented elsewhere.

Turning now to the analysis of the current-phase relation, we note that in the leading in $j_s$ order one has $f_L(x) \approx f_L^{c,1}$ inside the wire attached to the first superconducting terminal. Observing that (a) the kinetic equations are linear and (b) $f_L$ is an odd function of energy, we conclude that the function $ f_L^{c,1}$ can be expressed in terms of some linear combination of the functions $f_L^N(V_1),\ f_L^N(V_2)$ and $f_L^N(V_N)$. Since the value $|V_N|$ remains smaller than $V/2$, it follows immediately that $I_C^{(5)}(V)$ in the asymmetric five-terminals setup becomes suppressed at {\it higher voltages} as compared to that for the four-terminals setup. This observation is supported by the results of our numerical analysis displayed in Fig.~\ref{fig: Currents}.  We observe that the Josephson contribution evaluated for an asymmetric five-terminals setup survives up to the highest voltage values employed in the calculation (Fig.~\ref{fig: Currents}a). In this case for the same voltage range the Aharonov-Bohm current $I_m$ shows no sign of saturation, as it is indicated in Fig.~\ref{fig: Currents}b.

\begin{figure}
\centering
\includegraphics[width=1\linewidth]{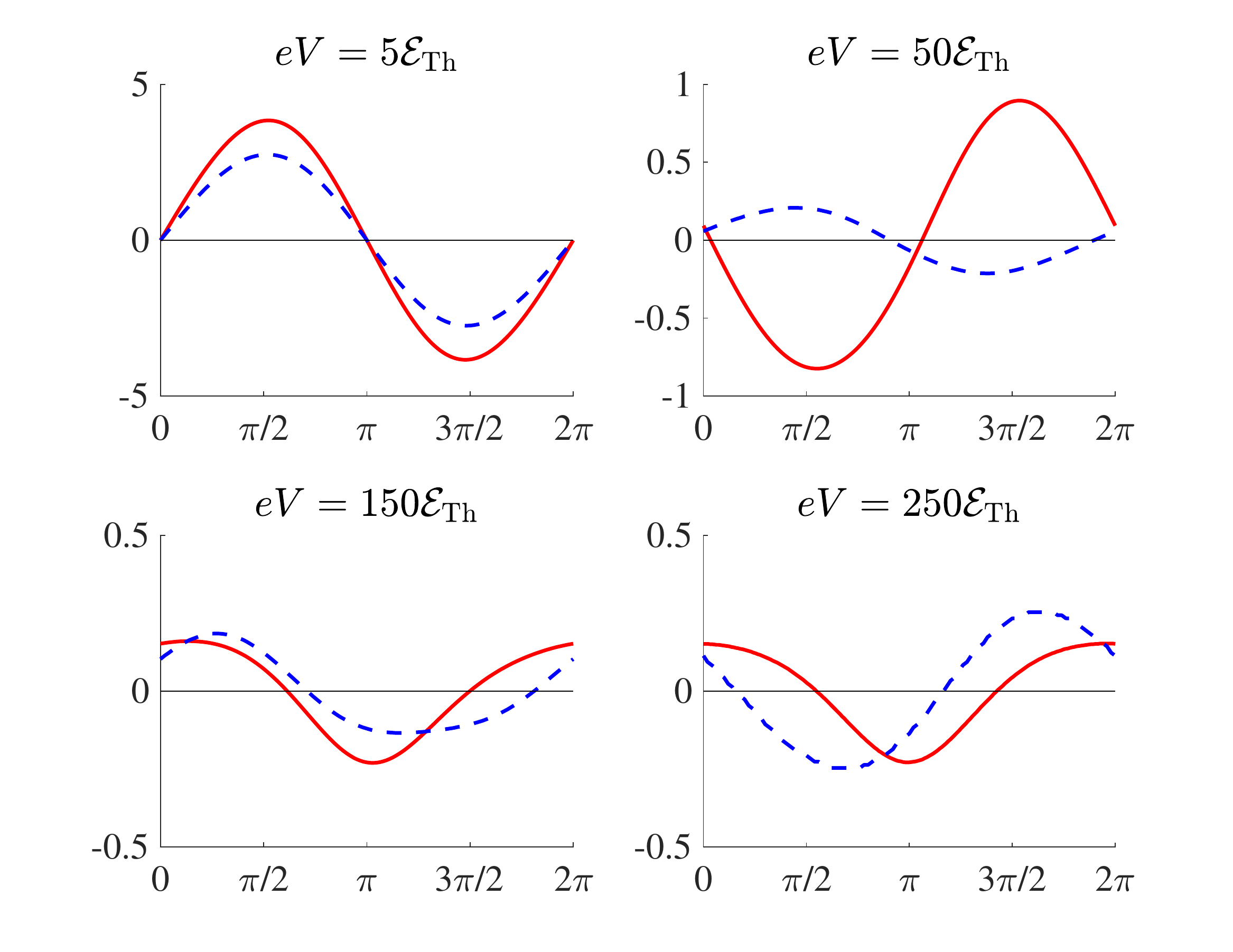}
\caption{The non-equilibrium current-phase relations for four- and asymmetric five-terminal setups (depicted respectively by solid and dashed curves)  at $T \to 0$ and different bias voltages $V$.}
\label{fig: Together}
\end{figure}

In Fig.~\ref{fig: Together} we further compare the full current-phase relations in both four- and five-terminal geometries at different bias voltages and $T \to 0$. In the four-terminals case -- in accordance with Eqs. (\ref{IJIAB})--(\ref{eq: I_m}) -- we observe a clear crossover from the odd-like behavior of $I_{\rm osc}(\phi)$ at lower voltages to the even-like one at higher values of $V$. By contrast, in five-terminal configurations the odd (Josephson-like) component remains dominant up to very high voltages. With increasing temperature, however, this component gets suppressed much stronger than $I_{\rm m}$, as it is illustrated in Fig.~\ref{fig: CMP_T}.

\section{Concluding remarks}
\label{sec: Conclusions}
In this work we investigated proximity-induced long-range quantum coherent effects in multi-terminal Andreev interferometers under non-equilibrium conditions.  We demonstrated that at low enough temperatures the current flowing between two superconducting terminals results from a non-trivial interplay between Josephson-like and Aharonov-Bohm-like effects. The corresponding contributions to the current $I_J$ and $I_{AB}$ are controlled both by the magnetic flux $\Phi$ threading the system and the external bias voltage $V$. As functions of the magnetic flux both currents $I_J$ and $I_{AB}$ exhibit coherent oscillations with the period $\Phi_0$ being respectively odd and even functions of $\Phi$. The magnitudes of these two current components demonstrate very different dependencies on both voltage bias and temperature, thus offering a unique opportunity to at will engineer the current-phase relation in Andreev interferometers.

The system topology is yet another important factor that may strongly affect its non-equilibrium behavior at low enough $T$. Here we demonstrated that by attaching an extra normal reservoir of electrons or just by changing the symmetry of our multi-terminal hybrid structure one can further modify both currents $I_J$ and $I_{AB}$ in a non-trivial manner. For instance, in the presence of the normal terminal N$_3$ (see Fig. 1)
some ``superconducting'' (i.e. phase-coherent) electrons propagating in the central normal wire get absorbed by this terminal being replaced by ``normal'' (i.e. insensitive to proximity-induced quantum coherence) electrons
from N$_3$. This process results in two (in part competing) effects: (i) quantum decoherence that yields partial {\it suppression} of both currents $I_J$ and $I_{AB}$ and (ii) modification in the electron distribution function that may produce significant {\it enhancement} of the Josephson component $I_J$ but has (almost) no extra effect on $I_{AB}$. We also discussed topology-dependent coherent oscillations of the voltage induced at the normal terminal isolated from the external leads.

Our predictions can be directly verified in modern experiments and may be used for designing
superconducting hybrid nanocircuits with controlled quantum properties.

\vspace{0.5cm}

\centerline{\bf Acknowledgements}

This work was supported in part by RFBR Grant No. 18-02-00586. P.E.D. acknowledges support by Skoltech NGP Program (Skoltech-MIT joint project).

\appendix
\section{}
\label{sec Appendix}

\label{appendix_1}

Consider first an SNS junction with a normal-metal wire of length $L$ connecting two superconducting terminals. At sufficiently high energies $\epsilon \gg {\cal E}_{\rm Th}$ the solution of the Usadel equation in the normal wire can be expressed as a superposition of the two independent anomalous propagators~\cite{ZZh}:
\begin{align}
{\cal F}_{12}(x) &= {\cal F}_{\rm SN}\Big(\frac{L}{2} + x\Big) e^{i\frac{\phi}{2}} + {\cal F}_{\rm SN}\Big(\frac{L}{2} - x\Big) e^{-i\frac{\phi}{2}},\label{eqn: F1}\\
{\cal F}_{21} (x) &= -{\cal F}_{\rm SN}\Big(\frac{L}{2} + x\Big) e^{-i\frac{\phi}{2}} - {\cal F}_{\rm SN}\Big(\frac{L}{2} - x\Big) e^{i\frac{\phi}{2}}\label{eqn: F2}
\end{align}
where
\begin{align}
{\cal F}_{\rm SN}(x) = -\frac{4 q (1 + q^2)}{(1-q^2)^2},\ q(x) = \frac{i}{1+\sqrt{2}} e^{x\sqrt{\frac{-2i\epsilon}{D}}}
\end{align}
and $x$ is the coordinate along the wire ($-L/2 \leq x \leq L/2$). Combining the above expressions with that for the spectral supercurrent
\begin{eqnarray}
j_s  &=& \frac{1}{4} \Tr \hat{\tau}_3 \Big( \hat{G}^R \partial_x \hat{G}^R - \hat{G}^A \partial_x \hat{G}^A\Big) =\notag{}\\
&=& \frac{1}{4} [ {\cal F}_{12} \partial_x {\cal F}_{21} -  {\cal F}_{21} \partial_x {\cal F}_{12} + c.c.]
\end{eqnarray}
we readily find
\begin{equation}
j_s = \frac{16}{3 + 2\sqrt{2}} \sin\phi \Big[ i \sqrt{\frac{-2i\epsilon}{D}} e^{\sqrt{\frac{-2i\epsilon}{D}}} + c.c.\Big].
\label{specjs}
\end{equation}
Having established this relation, we can now turn to the symmetric cross-like geometry~\cite{WSZ,Yip,Teun} with two superconducting and two normal terminals biased by the voltage $V$, cf. also Fig.~\ref{fig: geom} with $l_c=0$ and disconnected terminal N$_3$. Integrating Eq.~(\ref{specjs}) over energies and bearing in mind that at $T \to 0$ only states with energies $|\epsilon | \geq eV/2$ contribute to this integral, in the limit $eV \gg {\mathcal E}_{\rm Th}$ we obtain
\begin{equation}
I_C^{SNS} (V) \simeq \frac{32 (1+v^{-1})}{3+2\sqrt{2}} \frac{V}{R_L} e^{-v} \sin(v + v^{-1}),
\label{eq: I_C_SNS}
\end{equation}
where the parameter $v$ was defined above in Eq.~(\ref{IJIAB}).

\begin{figure}
\begin{minipage}[h]{1\linewidth}
\center{\includegraphics[width=1\linewidth]{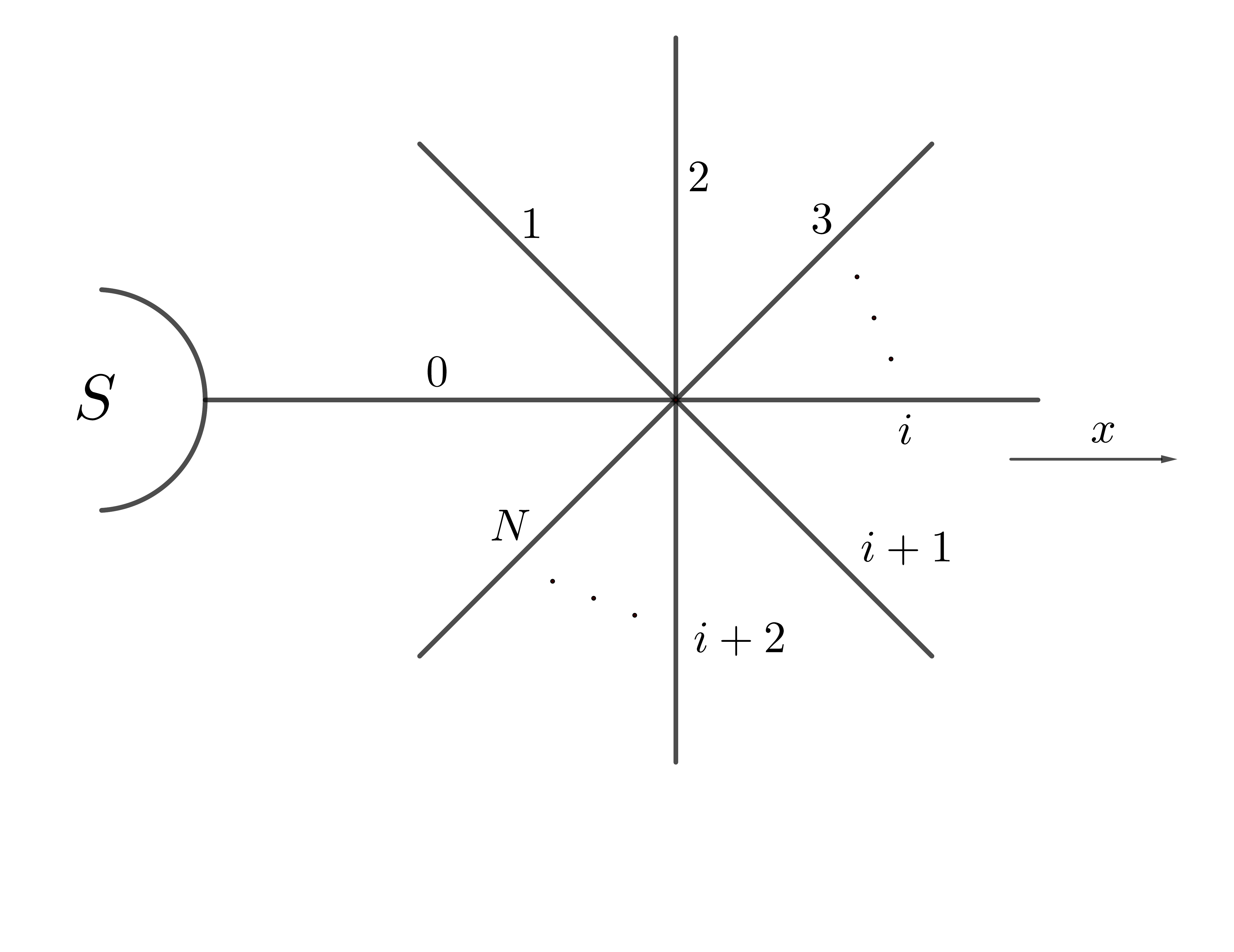}} \\a) \\
\end{minipage}
\vfill
\begin{minipage}[h]{1\linewidth}
\center{\includegraphics[width=1\linewidth]{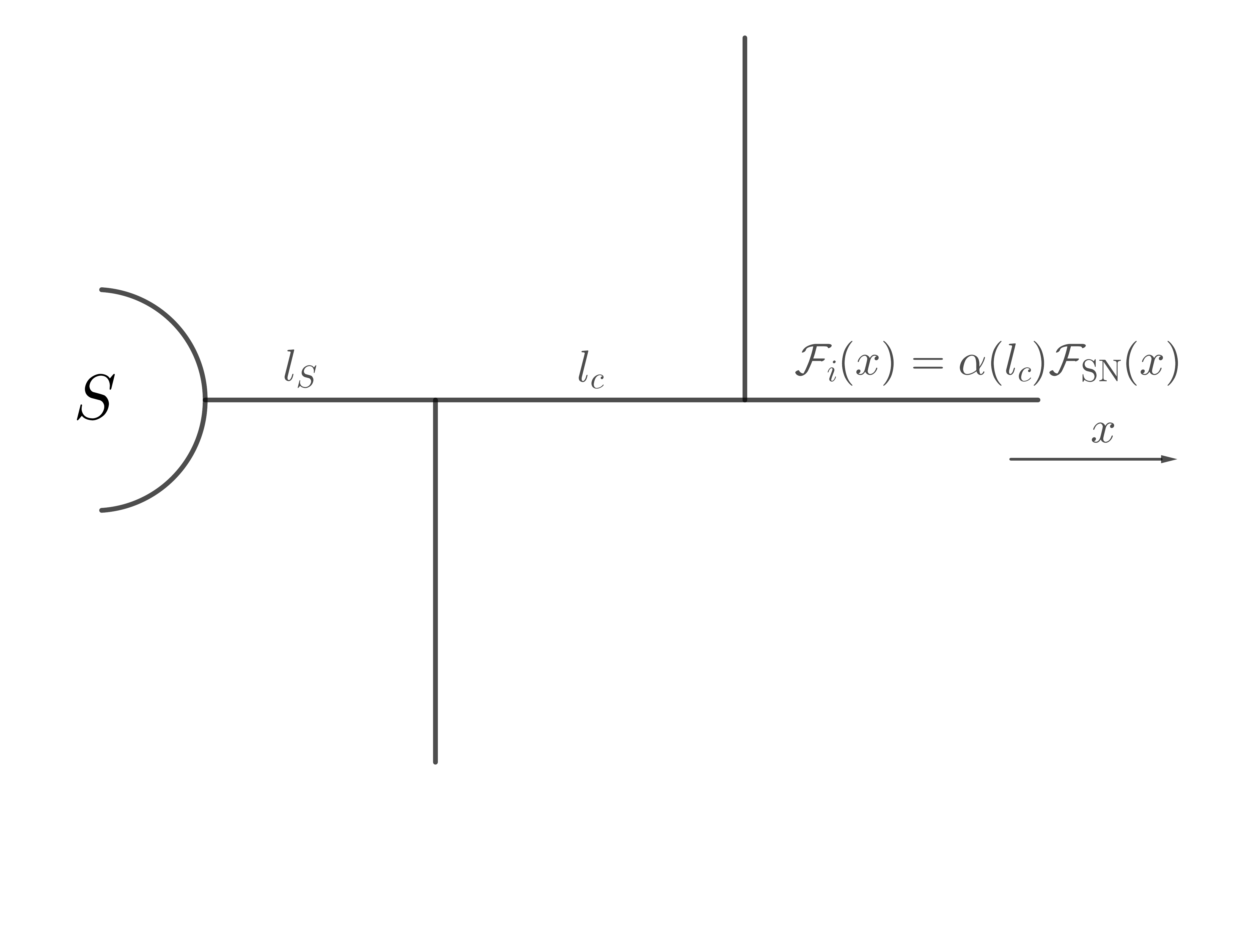}} \\b)
\end{minipage}
\caption{(a) A configuration with a superconducting terminal and attached normal wire(s) containing a branching point. (b)
The same with varying length $l_c$ between two branching points.}
\label{fig: Appendix}
\end{figure}

Note that this simple analysis totally neglects partial suppression of superconducting correlations inside the normal wires due to the presence of two extra normal terminals. The magnitude of this effect depends on the system topology and can be accounted for in a straightforward manner by formulating the matching conditions at the wire branching points. In a general form such conditions have already been discussed, e.g., in Refs.~\cite{Zaitsev,GWZ97}. In essence, they are equivalent to the current conservation law interpreted as a ``Kirchoff law for the Green functions'' -- see, e.g., Eq.~(29) in Ref.~\cite{GWZ97}.

Here we derive a simplified version of these matching conditions sufficient for our present purposes. To do so, we note that for energies $|\epsilon| \gg {\mathcal E}_{\rm Th}$ and at distances from a superconductor exceeding the length scale $\sim \sqrt{D/|\epsilon |}$ the Usadel equation can be linearized, thereby reducing to a simple wave-like equation:
\begin{equation}
D d^2{\cal F}/dx^2 + 2 i \epsilon {\cal F} = 0.\label{eqn: wave-eqn}
\end{equation}
As this equation is linear, the effect from the superconducting terminals can be treated independently, cf., e.g., Eqs.~(\ref{eqn: F1})--(\ref{eqn: F2}).

To this end let us consider a setup schematically depicted in Fig.~\ref{fig: Appendix}a. The setup consists of
one superconducting terminal attached to a normal wire which, in turn, is connected to $N$ other normal wires at some branching point located sufficiently far from the superconductor. By solving the wave-like equation~(\ref{eqn: wave-eqn}), we obtain the following matching conditions
\begin{equation}
{\cal F}_i(x) = \alpha {\cal F}_{\rm SN}(x),\ \alpha =2{\cal A}_0 / \Big(\sum\limits_{j = 0}^{N} {\cal A}_j \Big),\ i = 1,\dots, N, \label{eqn: branch}
\end{equation}
where ${\cal A}_i$ denotes the cross section of the $i$-{th} normal wire and $x$ is a coordinate along the corresponding wire. This condition implies that the anomalous propagator ${\cal F}_i(x)$ inside the $i$-th wire is suppressed by the factor $\alpha$ as compared to the propagator ${\cal F}_{\rm SN}(x)$ in the absence of the branching point ($N =1$).

Applying the matching condition (\ref{eqn: branch}) to the setup of Fig.~\ref{fig: geom} and bearing in mind that all wire cross sections are assumed to be equal ${\cal A}_i={\cal A}$,
for the anomalous propagator ${\cal F}_{12}(x)$ inside the wire $l_c$ we obtain
\begin{equation}
{\cal F}_{12}(x) = \alpha {\cal F}_{\rm SN}\Big(\frac{L}{2} + x\Big) e^{i\frac{\phi}{2}} + \tilde\alpha{\cal F}_{\rm SN}\Big(\frac{L}{2} - x\Big) e^{-i\frac{\phi}{2}},\label{eqn: F3}
\end{equation}
with $\alpha=\tilde\alpha =2/3$ in the absence of the terminal N$_3$ (four-terminal configuration) and
$\alpha =2/3$, $\tilde\alpha = 4/9$ (or vice versa depending on whether $x$ is located to the left or to the right with respect to the wire $l_{N,3}$) for the five-terminal setup in Fig.~\ref{fig: geom}. Accordingly, in these two cases the currents $I_C^{(4)} (V)$ and $I_{C,0}^{(5)} (V)$ get reduced respectively by the factors $4/9$ and $8/27$ compared to $I_C^{SNS} (V)$ in Eq.~(\ref{eq: I_C_SNS}).

Note that, strictly speaking, the latter results for the prefactors  $\alpha$ and $\tilde\alpha$ apply only provided different branching points in the setup of Fig.~\ref{fig: geom} are located sufficiently far from each other. In order to address a more general situation let us consider the configuration depicted in Fig.~\ref{fig: Appendix}b with the wire length $l_c$ varying from zero to large values.  According to Eq.~(\ref{eqn: branch}),
for $l_c \to 0$ we have $\alpha (0) = 1/2$ (we again assume that all wires have the same cross section), while in the limit of large $l_c$ one finds $\alpha = (2/3)^2 = 4/9$. Making use of the linearity of Eq.~(\ref{eqn: wave-eqn}) and the conservation of the spectral currents, we recover the complete dependence $\alpha(l_c)$:
\begin{equation}
\alpha(l_c) \approx  4\left( 9 - \exp\left(2 l_c \sqrt{\frac{-2i\epsilon}{D}}\right)  \right)^{-1}.
\end{equation}
This result demonstrates that in a general case the function $\alpha(l_c)$ depends on energy.

Bearing in mind the above matching conditions, for the setup in Fig.~\ref{fig: geom} we obtain
\begin{eqnarray}
I_C^{(4)} (V) &\simeq& I_C^{SNS} (V) \begin{cases} \frac{4}{9} & \mbox{if } l_c^2 \gg D/(eV) \\
\frac{1}{2} & \mbox{if } l_c^2 \ll D/(eV)  \end{cases},\\
I_{C,0}^{(5)} (V) &\simeq& I_C^{SNS} (V) \begin{cases} \frac{8}{27} & \mbox{if } l_c^2 \gg D/(eV) \\
\frac{2}{5} & \mbox{if } l_c^2 \ll D/(eV) \end{cases}.
\end{eqnarray}

\end{document}